\documentclass[12pt]{iopart}

\usepackage{graphicx}
\usepackage{enumitem}
\usepackage{url} 

\expandafter\let\csname equation*\endcsname\relax
\expandafter\let\csname endequation*\endcsname\relax
\usepackage{amsmath}

\begin{document}

\title[Friction Induced Energy Gain]{Friction Induced Energy Gain}

\author{Diego M Fieguth}

\address{State reaserch center OPTIMAS and Fachbereich Physik, Rheinland-Pf\"alzische Technische Universit\"at Kaiserslautern-Landau,D-67663 Kaiserslautern, Germany}
\ead{fieguth@rhrk.uni-kl.de}
\vspace{10pt}
\begin{indented}
\item[]December 2024
\end{indented}

\begin{abstract}
In this work we show how friction enables a non-linear energy transfer in a slow-fast Hamiltonian system. We first introduce a paradigmatic system consisting of a weakly coupled fast and slow oscillator that gives rise to a non-linear resonance. We state known Assertions about this system and the conservation of energy in the slow variables.
We reexamine the system with friction in the slow degrees and identify two effects that lead to an energy transfer from the fast degrees of freedom to the slow degrees.
We then replace the slow harmonic oscillator with an asymmetric double well system.
Only with friction turned on the particle can move from lower energy to a high energy stable
equilibrium point.
\end{abstract}

%
%
\submitto{\NL}
%
%
%

\section{Introduction}

A recipe for practical energy storage becomes apparent when looking at its units, $\mathrm{kg\,m^2}/\mathrm{s}^2$. To densely store energy in a finite volume we have to make the depot heavy or fast. There is a practical limit on the mass density which leaves a fast time scale as solution for densely packing energy. The reason for the energy contained in fossil fuels are the fast atomic and molecular timescales.

Useful work, like moving a car for example, acts on a much slower timescale. This time scale hierarchy and the energy transfer between fast and slow degrees of freedom are important problems \cite{Daemon,QDaemon,kolomeisky2007molecular} and have been recognized to play an important role in enzyme dynamics \cite{Henzler2007}.

Systems with different time scales usually act as if decoupled, but there are non-linear resonances, where such an energy transfer can happen. Non-linear resonances are known to play an important dynamical role in classical molecule dynamics of the classical three body Coulomb problem \cite{itin2003resonant}.
Important theorems on non-linear resonances, like capture and scattering and the quasi-probabilistic splitting of trajectories ar long known classically \cite{neishtadt2006destruction,Neishtadt2014,Henrard,Artemyev,Neishtadt1996,Neishtadt2000,Neishtadt2017} and recently gained some quantum-mechanical/ semi-classical analogues \cite{kuksin2013quantum,stabel2022dynamical}.

In this work we set up a paradigmatic slow-fast system to demonstrate the unintuitive result that friction can \emph{assist} this energy transfer. 
First  we show numerical simulations of a system consisting of two non-linearly coupled harmonic oscillators (one fast, one slow) and state Assertions for the conservative system. We then  show the same system with friction in the slow degrees of freedom. We discuss how the Assertions change and how this allows a transfer of energy from the fast to the slow degrees of freedom.

Then we show how the effects introduced with friction can induce a transition across a barrier in an asymmetric double well system from the lower energy well to the higher energy well.
\section{Slow-fast Hamiltonian system}
We start with two harmonic oscillators. The slow time scale is $\varepsilon\ll1$ and the Hamiltonian is\begin{equation}
    H_0(\varepsilon^{-1}Q,P,q,p)=\frac{P^2+\Omega^2 Q^2}{2}+\frac{p^2+\omega^2q^2}{2}. \label{eq:H0}
\end{equation}

The time evolutions generated by $H_0$ for the two pairs of canonical coordinates is decoupled and trajectories are lines of conserved energy, circles in the phase space. We label the energy of the slow variables $E=\frac{P^2+\Omega^2 Q^2}{2}$. Transferring energy from fast to slow degrees of freedom means increasing $E$.

Next we introduce the weak coupling \begin{equation}
    \varepsilon^{r}H_1(\varepsilon^{-1}Q,q)=-\varepsilon^r q \cos(k\varepsilon^{-1}Q). \label{eq:H1}
\end{equation} The exponent $r<1$ so we have an additional separation $\varepsilon\ll\varepsilon^r\ll1$. This scale separation is not strictly necessary (e.g. in \cite{Neishtadt1996,Neishtadt2005} $r=1$), but it enhances the effect we want to present.

\begin{figure}
    \centering
    \includegraphics[width=0.49\linewidth]{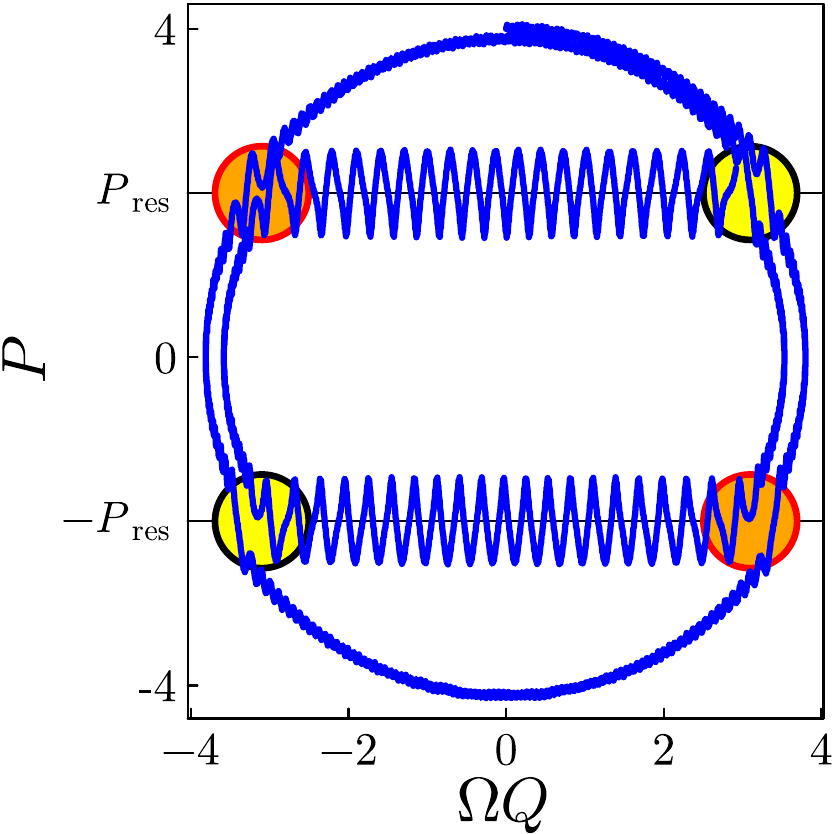}
    \includegraphics[width=0.49\linewidth]{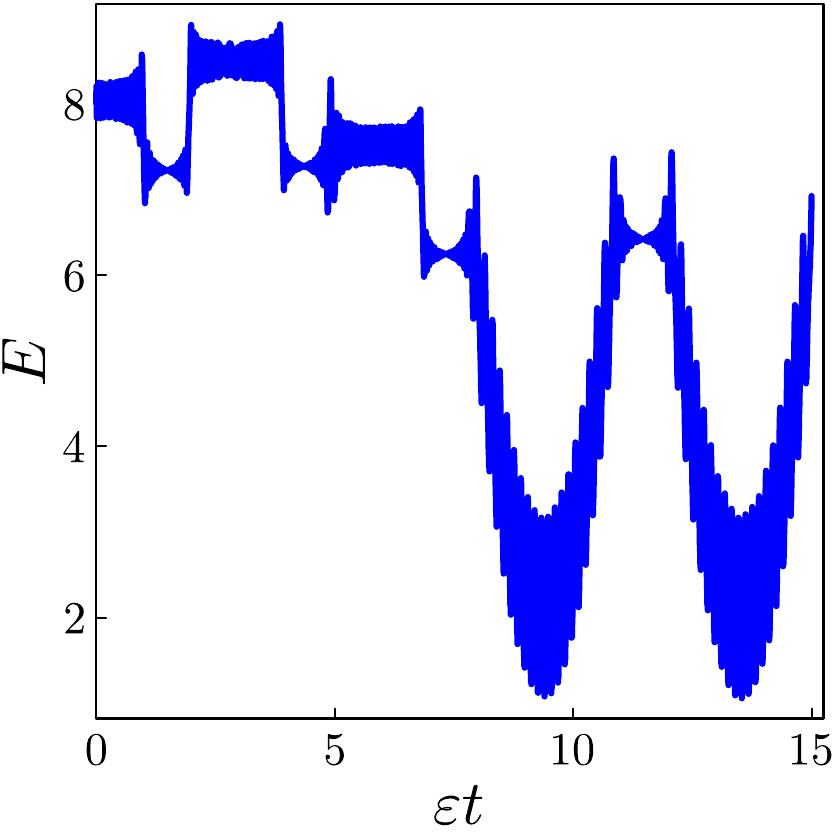}
    \caption{\textbf{Left panel:} Numerical solution of the slow canonical variables' equations of motion using $H_0+\varepsilon^r H_1$ (see \eqref{eq:H0},\eqref{eq:H1}). Intersections wit the resonant surface at $P=\pm P_\mathrm{res}$ are indicated in orange/red for regions where capture into resonance is possible, and yellow/black where escape from the resonant surface happens.\textbf{Right panel:} Energy $E$ stored in the slow variables. }
    \label{fig:frictionless}
\end{figure}

The time evolution generated by $H_0+\varepsilon^r  H_1$ can bee seen in Figure \ref{fig:frictionless}.
We see that non-linear resonances \cite{Chirikov,neishtadt2006destruction} are introduced to the system when looking at its time evolution. Away from the resonances, time evolution is as if decoupled.

The system can skip past the so called resonant surface at $P=\pm P_\mathrm{res}$ (scattering on a resonance) or can travel along the resonant surface (capture into resonance). The intersections of the line of constant energy in the phase space of the slow variables are marked in color. Orange markers indicate that the energy decreases along the resonant surface, yellow markers indicate that energy increases along the resonant surface. Scattering on the resonance happens in both the orange and the yellow regions in the left panel of Figure \ref{fig:frictionless}. Capture into resonance happens in the orange region and the subsequent escape from the resonance happens in the yellow region.
In the right panel of Figure \ref{fig:frictionless} we see $E$ over slow time. Every time the system skips past the resonant surface, (scattering) there is a small jump in $E$ (see the jumps before $\varepsilon t=5$ for example). If the system is captured into a resonance there is an initial loss in $E$ followed by a subsequent gain, ending up at approximately the same energy the system started with, which can be seen as the two dips in $E$.

To obtain the theoretical values of these resonances one can follow standard procedures \cite{neishtadt2006destruction,Neishtadt2017,artemyev2017probabilistic}. Transforming the fast system to action angle variables and using trigonometric identities one obtains the resonant angles $\rho_\pm=\alpha \pm k\varepsilon^{-1}Q$.
The points where $\dot{\rho}_\pm=0$ corresponds to $P=\pm \frac{\omega}{k}=:\pm P_\mathrm{res}$, the resonant surfaces. 

The ``decision'' between scattering or capturing on a resonant surface is quasi-probabilistic. It can be analyzed in a transformed pendulum-like system \cite{Neishtadt1996,eichmann2018engineering,Thesing2017} where glued adiabatic solutions \cite{Neishtadt1996} determine the trajectory. At the bifurcation there is a certain probability of capture given by the Kruskal-Neishtadt-Henrard formula \cite{eichmann2018engineering,Neishtadt2017,Henrard}.

Analyzing the Hamiltonian system using glued adiabatic solutions, yields trajectories that can only temporarily change their energy temporarily (in zeroth order in $\varepsilon^r$).

The important emergent properties of the evolution of the slow variables in the conservative system, shown in Figure \ref{fig:frictionless} are described by the following Assertions:\begin{enumerate}[label=(\alph*)]
    \item\label{a} time evolution away from the resonant surfaces is as if unperturbed, $E\approx \mathrm{const}$ 
    \item \label{b} behavior at the resonant surfaces is quasi-probabilistic with probabilities described by the KNH theorem
    \item \label{c} if travel along the resonant surface increases the energy $E$ (yellow markers) the probability of capture is zero
    \item \label{d} if travel along the resonant surface lowers the energy $E$ (orange markers) the probability of capture can be larger than zero
    \item \label{e} captured trajectories leave the resonant surface at approximately the same energy $E$
\end{enumerate}

The resonance is \emph{homoenergetic}. It does not change the energy $E$ by a large amount.

\section{Behaviour  change with friction}
Having seen that the conservative Hamiltonian system cannot change the energy of the slow variables (Assertion \ref{e}), we now introduce friction by changing the EOM of the slow momentum\begin{equation}
    \dot{P}=-\left( \frac{\partial H   }{\partial Q} +\gamma P  \right).
\end{equation}
\begin{figure}
    \centering
    \includegraphics[width=0.49\linewidth]{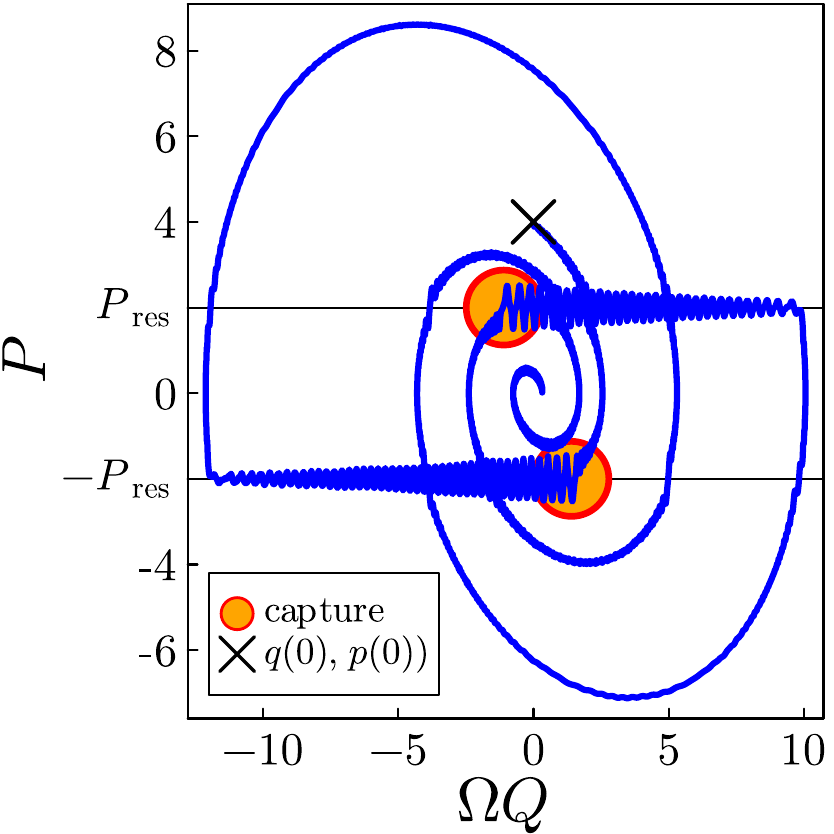}
    \includegraphics[width=0.49\linewidth]{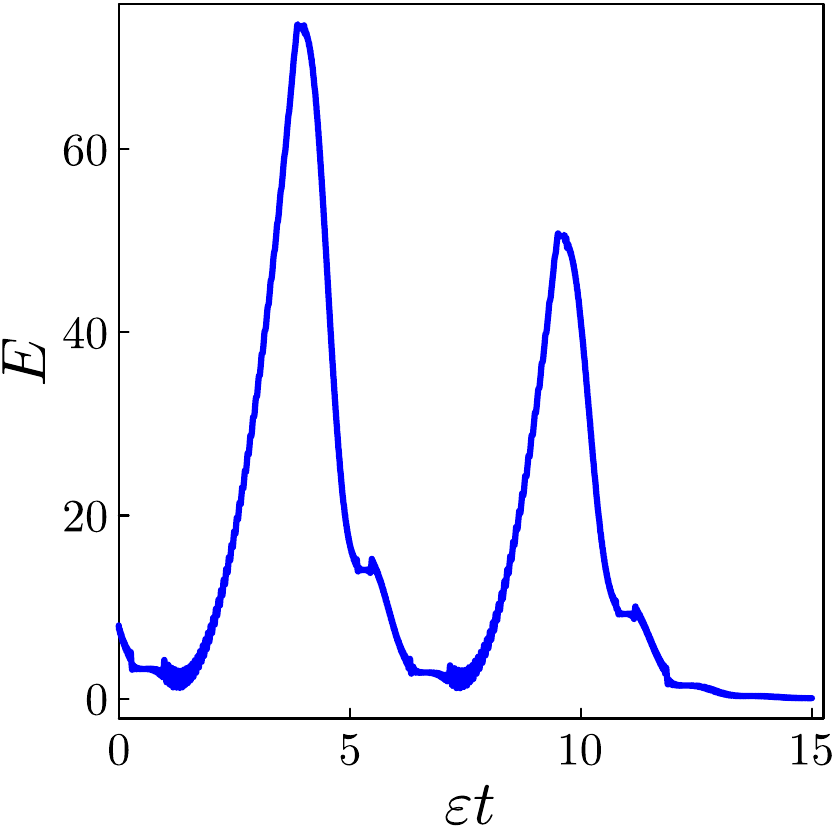}
    \caption{Two numerical solutions with friction with different initial conditions.\textbf{Left panel:} The orange area plays the same role as in Figure \ref{fig:frictionless} \textbf{Right panel:} Energy $E$ stored in the slow variables. }
    \label{fig:friction_longerstay}
\end{figure}

\begin{figure}
    \centering
    \includegraphics[width=0.49\linewidth]{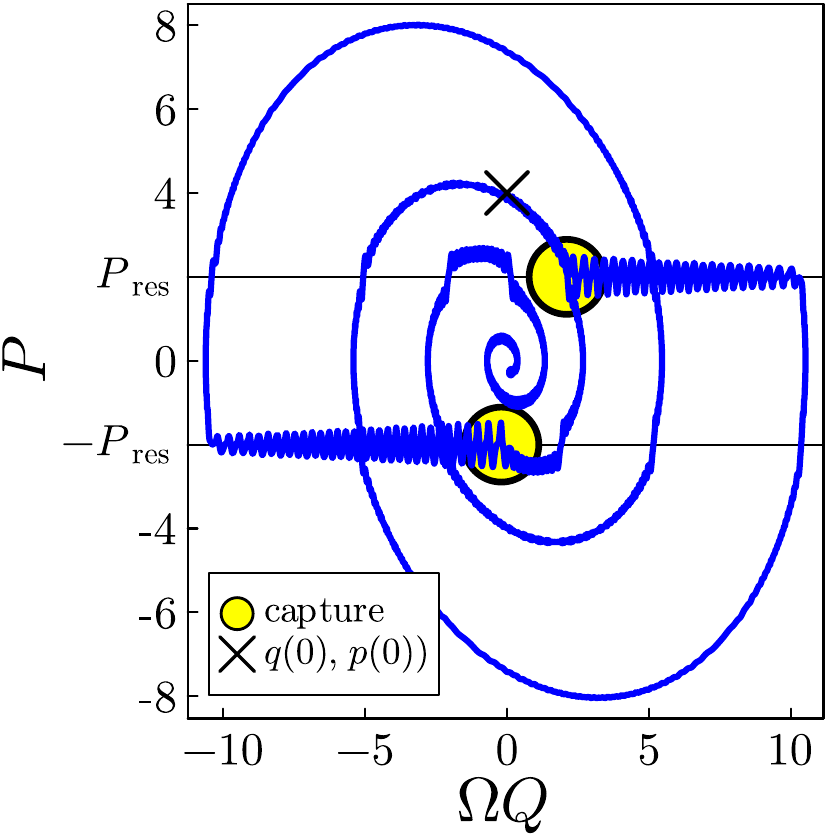}
    \includegraphics[width=0.49\linewidth]{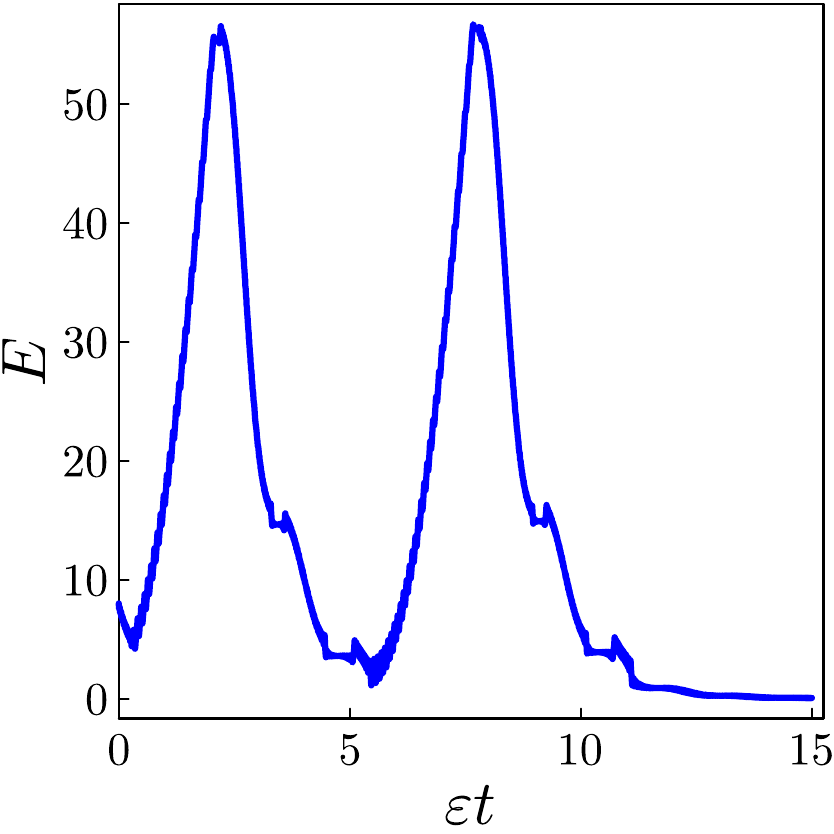}
    \caption{Two numerical solutions with friction with different initial conditions.\textbf{Left panel:} The orange area plays the same role as in Figure \ref{fig:frictionless} \textbf{Right panel:} Energy $E$ stored in the slow variables. }
    \label{fig:friction_heterocap}
\end{figure}

The simulations that show the effects are shown in Figures \ref{fig:friction_heterocap} and \ref{fig:friction_longerstay}. The two Figures look deceptively similar but they highlight different effects which we explain in the text below.

To better understand the effects we marked regions with the same colors as before in Figure \ref{fig:frictionless}, regions with energy decreasing along the resonant surface show orange markers.

First, in Figure \ref{fig:friction_longerstay} we see a \emph{delayed escape} from the resonance. Capture happens in the orange regions, just like in the frictionless case (Assertion \ref{d} does not change). In the case with friction the system does \emph{not} escape from the resonant surface at approximately the same energy but leaves much later at a higher energy in the slow coordinates. The energy transfer from fast to slow coordinates is significant, as can be seen in the right panel of Figure \ref{fig:friction_longerstay}. 
One can understand this in the approximated pendulum-like system \cite{Neishtadt1996,eichmann2018engineering,Thesing2017} in the vicinity of the resonant surface. The averaged equations for the action variable in this space is an exponential decay. In the slow coordinates this decay manifests itself in a decaying amplitude of the oscillation around the resonant surface. The separatrix in the reduced phase space has to shrink to a smaller size for the trajectory to exit opposed to the case where action stays constant.

The second effect can be seen in Figure \ref{fig:friction_heterocap}. The key difference to the conservative system is that capture happens in a region, where capture is impossible without friction. For the system with friction Assertion \ref{c} does no longer hold. This is because of an increased Kruskal-Neishtadt-Henrard probability. The mathematics of this is discussed in \cite{fieguth2023open}. In a region in the $Q-P$ phase space where without friction capture is impossible, the quasi-probability of capture can now can be greater than zero.  Looking at the numerator of the probability formula \cite{fieguth2023open} for this system, with separatrix area $A_\mathrm{sep}(t)$, we see that it is \begin{equation}
    \dot{A}_\mathrm{sep}+\gamma A_\mathrm{sep}.
\end{equation} The conservative case is for $\gamma=0$. There capture is impossible if the area of the separatrix is shrinking. For $\gamma\neq0$ this restriction can be overcome by the term $\gamma A_\mathrm{sep}$, which can lead to a finite probability if   $\dot{A}_\mathrm{sep}+\gamma A_\mathrm{sep}>0$.

We have seen that friction can massively benefit the energy gain for the slow coordinates by the two effects of delayed escape and increased capture probability.

\section{Friction induced transition}
With friction in the slow coordinates and a harmonic potential every energy gain is eventually lost to friction, when the particle comes to rest at the stable equilibrium point. To have a lasting energy gain we take an asymmetric double well potential. The right well is at lower energy than the left well. We now show that for a particle starting in the lower energy well can gain enough energy to overcome the barrier using a resonance. This is only possible with friction present. Without friction travel along the resonant surface approximately conserves $E$ (Assertion \ref{e}). With friction we have a \emph{friction induced energy gain}.

\begin{figure}
    \centering
    \includegraphics[width=0.49\linewidth]{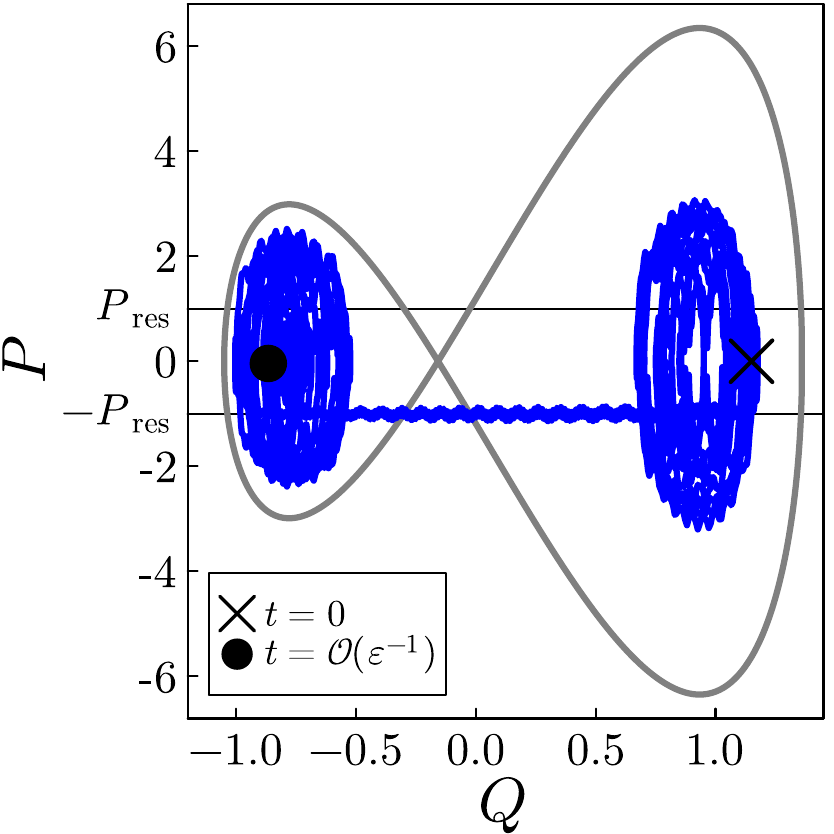}
    \includegraphics[width=0.49\linewidth]{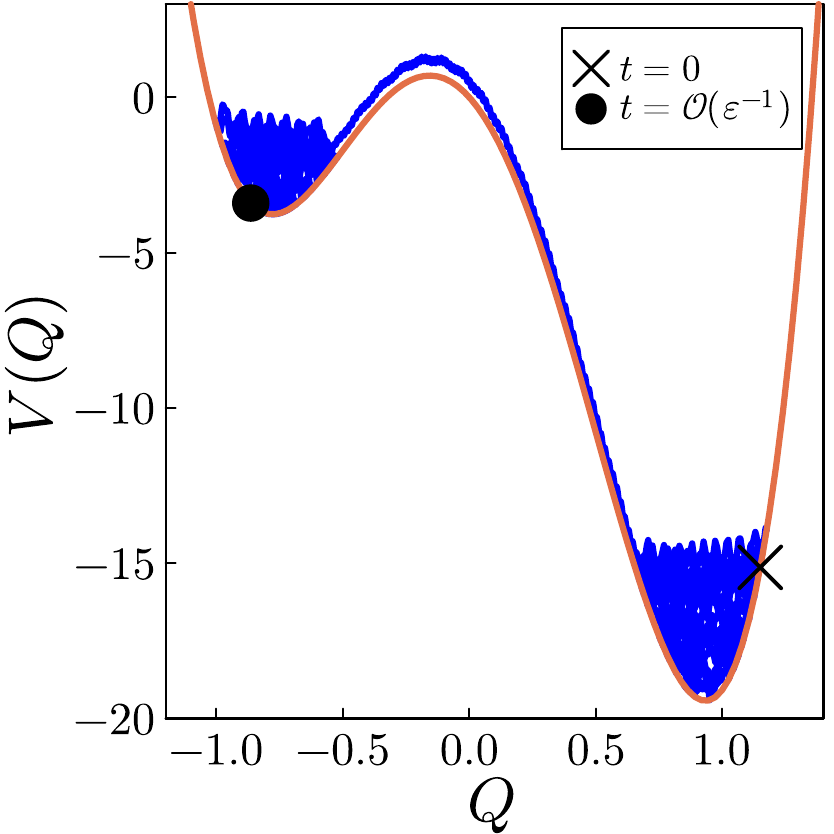}
    \caption{Numerical solutions with friction in a double well potential.\textbf{Left panel:} Trajectory of the system in the slow phase space \textbf{Right panel:} Energy $E$ of the slow variables. }
    \label{fig:fiet}
\end{figure}

The friction induced energy gain can be seen in Figure \ref{fig:fiet}. In the left panel we can see the slow phase space. The state starts with low energy in the deeper right well. Then it gets captured into resonance and gets transported along the resonant surface to the left well, where it escaped the resonance. 
The energy landscape is shown in the right panel of \ref{fig:fiet}. There we see that the energy gain is not a small effect and that the energy gain is permanent.

\section{Conclusion}

We have seen that the important task of transferring energy from fast to slow degrees of freedom can be \emph{enhanced} by friction in the slow variables. There are two mechanisms that improve the energy transfer from a resonance: delayed exit from a resonance and enhanced capture into the resonance. We have seen that the system can even overcome an energy barrier by using the energy stored in the fast degree of freedom, which is inaccessable without friction. 

\ack
The author thanks James R. Anglin for the valuable discussion on the topic and acknowledges the support from State Research Center OPTIMAS and the Deutsche Forschungsgemeinschaft (DFG) through SFB/TR185 (OSCAR), Project No. 277625399.

\bibliographystyle{vancouver}
\bibliography{citation}

\providecommand{\noopsort}[1]{}\providecommand{\singleletter}[1]{#1}%
\begin{thebibliography}{10}

\bibitem{Daemon}
Gilz L, Thesing E, Anglin JR.
\newblock Hamiltonian analogs of combustion engines: A systematic exception to adiabatic decoupling.
\newblock Phys Rev E. 2016 Oct;94:042127.
\newblock Available from: \url{https://link.aps.org/doi/10.1103/PhysRevE.94.042127}.

\bibitem{QDaemon}
Thesing EP, Gilz L, Anglin JR.
\newblock Quantum Hamiltonian daemons: Unitary analogs of combustion engines.
\newblock Phys Rev E. 2017 Jul;96:012119.
\newblock Available from: \url{https://link.aps.org/doi/10.1103/PhysRevE.96.012119}.

\bibitem{kolomeisky2007molecular}
Kolomeisky AB, Fisher ME.
\newblock Molecular motors: a theorist's perspective.
\newblock Annu Rev Phys Chem. 2007;58(1):675-95.

\bibitem{Henzler2007}
Henzler-Wildman KA, Lei M, Thai V, Kerns SJ, Karplus M, Kern D.
\newblock A hierarchy of timescales in protein dynamics is linked to enzyme catalysis.
\newblock Nature. 2007;450(7171):913-6.

\bibitem{itin2003resonant}
Itin A.
\newblock Resonant phenomena in classical dynamics of three-body Coulomb systems.
\newblock Physical Review E. 2003;67(2):026601.

\bibitem{neishtadt2006destruction}
Neishtadt A, Vasiliev A.
\newblock Destruction of adiabatic invariance at resonances in slow--fast Hamiltonian systems.
\newblock Nuclear Instruments and Methods in Physics Research Section A: Accelerators, Spectrometers, Detectors and Associated Equipment. 2006;561(2):158-65.

\bibitem{Neishtadt2014}
Neishtadt AI.
\newblock Averaging, passage through resonances, and capture into resonance in two-frequency systems.
\newblock Russian Mathematical Surveys. 2014;69(5):771.

\bibitem{Henrard}
Henrard J.
\newblock {Capture into resonance: An extension of the use of adiabatic invariants}.
\newblock Celestial mechanics. 1982;27(1):3-22.

\bibitem{Artemyev}
Artemyev AV, Neishtadt AI, Vainchtein DL, Vasiliev AA, Vasko IY, Zelenyi LM.
\newblock Trapping (capture) into resonance and scattering on resonance: Summary of results for space plasma systems.
\newblock Communications in Nonlinear Science and Numerical Simulation. 2018;65:111-60.
\newblock Available from: \url{https://www.sciencedirect.com/science/article/pii/S1007570418301485}.

\bibitem{Neishtadt1996}
Neishtadt A.
\newblock Scattering by resonances.
\newblock Celestial Mechanics and Dynamical Astronomy. 1996;65:1-20.

\bibitem{Neishtadt2000}
Neishtadt AI.
\newblock On the accuracy of persistence of adiabatic invariant in single-frequency systems.
\newblock Regular and chaotic dynamics. 2000;5(2):213-8.

\bibitem{Neishtadt2017}
Neishtadt A.
\newblock Averaging method for systems with separatrix crossing.
\newblock Nonlinearity. 2017;30(7):2871.

\bibitem{kuksin2013quantum}
Kuksin SB, Neishtadt AI.
\newblock On quantum averaging, quantum KAM, and quantum diffusion.
\newblock Russian Mathematical Surveys. 2013;68(2):335.

\bibitem{stabel2022dynamical}
Stabel P, Anglin JR.
\newblock Dynamical change under slowly changing conditions: the quantum Kruskal--Neishtadt--Henrard theorem.
\newblock New Journal of Physics. 2022;24(11):113052.

\bibitem{Neishtadt2005}
Neishtadt A.
\newblock Probability phenomena in perturbed dynamical systems.
\newblock In: Mechanics of the 21st Century: Proceedings of the 21st International Congress of Theoretical and Applied Mechanics, Warsaw, Poland, 15--21 August 2004. Springer; 2005. p. 241-61.

\bibitem{Chirikov}
Chirikov BV.
\newblock Research concerning the theory of non-linear resonance and stochasticity.
\newblock CM-P00100691; 1971.

\bibitem{artemyev2017probabilistic}
Artemyev A, Neishtadt A, Vasiliev A, Mourenas D.
\newblock Probabilistic approach to nonlinear wave-particle resonant interaction.
\newblock Physical Review E. 2017;95(2):023204.

\bibitem{eichmann2018engineering}
Eichmann T, Thesing EP, Anglin JR.
\newblock Engineering separatrix volume as a control technique for dynamical transitions.
\newblock Physical Review E. 2018;98(5):052216.

\bibitem{Thesing2017}
Gilz L, Thesing E, Anglin JR.
\newblock Hamiltonian analogs of combustion engines: A systematic exception to adiabatic decoupling.
\newblock Physical Review E. 2016;94(4):042127.

\bibitem{fieguth2023open}
Fieguth DM, Anglin JR.
\newblock Open system control of dynamical transitions under the generalized Kruskal-Neishtadt-Henrard theorem.
\newblock Physical Review E. 2023;107(3):034209.

\end{thebibliography}

\end{document}